%% file: main.tex
\documentclass[conference]{IEEEtran}
\usepackage{fancyhdr}

\usepackage{xcolor,soul,framed} 
\usepackage{siunitx}
\usepackage{gensymb}
\usepackage{tabularx}

\colorlet{shadecolor}{yellow}

\usepackage[pdftex]{graphicx}
\graphicspath{{../pdf/}{../jpeg/}}
\DeclareGraphicsExtensions{.pdf,.jpeg,.png}

\usepackage[cmex10]{amsmath}

\usepackage{array}
\usepackage{mdwmath}
\usepackage{mdwtab}
\usepackage{eqparbox}
\usepackage{url}
\usepackage{booktabs}

\usepackage{graphicx}
\usepackage{float}
\usepackage{textcomp}
\usepackage{xspace}
\usepackage{xr}

\usepackage{amsmath}
\let\labelindent\relax
\usepackage{enumitem}
\usepackage{amssymb}
\usepackage[super]{nth}
\usepackage{multirow}
\usepackage{diagbox}
\usepackage{csquotes}
\usepackage[nolist]{acronym}
\usepackage{epsfig}
\usepackage{cleveref}

\usepackage{array}

\usepackage[pscoord]{eso-pic}

\begin{document}

\begin{acronym}
  \acro{de}[DE]{distance estimation}
  \acro{dess}[DESS]{DE relying on signal strength}
  \acro{tof}[ToF]{time-of-flight}
  \acro{rss}[RSS]{received signal strength}
  \acro{uwb}[UWB]{ultra-wideband}
  \acro{ml}[ML]{machine learning}
  \acro{mae}[MAE]{mean absolute error}
  \acro{rmse}[RMSE]{root mean square error}
  \acro{cots}[COTS]{commercial off-the-shelf}
  \acro{los}[LOS]{line-of-sight}
  \acro{cir}[CIR]{channel impulse response}
  \acro{agc}[AGC]{automatic gain control}
  \acro{mi}[MI]{multipath interference}
\end{acronym}

\title{Improving Signal-Strength-based Distance Estimation in UWB Transceivers
}

\author{
    \IEEEauthorblockN{
    Leo Botler\IEEEauthorrefmark{1}\IEEEauthorrefmark{2},
    Konrad Diwold\IEEEauthorrefmark{1}\IEEEauthorrefmark{3} and
    Kay R\"omer\IEEEauthorrefmark{1}\IEEEauthorrefmark{2}}
    \IEEEauthorblockA{
    \small \IEEEauthorrefmark{1}Institute of Technical Informatics \\ Graz University of Technology, Austria \\}
    \IEEEauthorblockA{\IEEEauthorrefmark{2}\{leo.happbotler, kdiwold, roemer\}@tugraz.at}
    \IEEEauthorblockA{\IEEEauthorrefmark{3}Pro2Future GmbH, Inffeldgasse 25F, 8010 Graz, Austria, konrad.diwold@pro2future.at}
    }

\maketitle

\thispagestyle{fancy}
  
\begin{abstract}
\Ac{uwb} technology has become very popular for indoor positioning and \ac{de} systems due to its decimeter-level accuracy achieved when using time-of-flight-based techniques. Techniques for \ac{dess} received less attention. As a consequence, existing benchmarks consist of simple channel characterizations rather than on methods aiming to increase accuracy. Further development in \ac{dess} may enable lower-cost transceivers to applications which can afford lower accuracies than those based on time-of-flight. Moreover, it is a fundamental building block used by a recently proposed approach that can enable security against cyberattacks to \ac{de} which could not be avoided using only time-of-flight-based techniques. 
In this paper, we evaluate the suitability of several machine-learning models trained in different real-world environments to increase \ac{uwb}-based \ac{dess} accuracy. Additionally, aiming implementation in \ac{cots} transceivers, we propose and evaluate an approach to resolve ambiguities comprising \ac{dess} in these devices.  Our results show that the proposed \ac{de} approaches have sub-decimeter accuracy when testing the models in the same environment and positions in which they have been trained, and achieved an average MAE of \SI{24}{\cm} when tested in a different environment. 3 datasets obtained from our experiments are made publicly available.    
\end{abstract}


\input{./sections/1-introduction}

\input{./sections/2-approach-and-methods}
\input{./sections/3-experiments}
\input{./sections/4-analysis-and-results}

\input{./sections/5-related-work}
\input{./sections/6-conclusion}
\input{./sections/8-acknowledgments}


\bibliographystyle{IEEEtran}
\bibliography{IEEEabrv}

\end{document}

%% file: sections/1-introduction.tex

\section{INTRODUCTION}
\label{sec:introduction}

Accurate \ac{de} is an enabler for several applications, including Passive Keyless Entry and Start (PKES), and Indoor Positioning Systems. \Ac{de} can currently be achieved by \ac{cots} \ac{uwb} transceivers, which enable accurate \ac{de} using \ac{tof} with a sub-decimeter accuracy~\cite{dw1000-datasheet}.

\ac{de} approaches relying on \ac{uwb}'s signal's strength have been less explored. Typical approaches make use of a single feature of the signal, namely the first path amplitude of the signal reaching the receiver, and characterize the error based on the standard deviation of this feature.

In general, experiments making use of laboratory equipment differ from those using \ac{cots} in the sense that they 1) afford a higher and more stable sampling frequency, 2) do not use an \ac{agc} stage in the receiver - which will be shown in Section~\ref{sec:analysis-and-results} to be critical - and, 3) make use of wider bandwidths than those allowed by standards, which directly impacts the accuracy of distance estimations.

In this paper, we improve the state-of-the-art accuracy of \ac{uwb}-based \ac{dess} by using \ac{ml} regressors. To our knowledge, this is the first time that \ac{ml} is applied to \ac{uwb}-based \ac{dess}. We do not aim to achieve more accurate estimations than those achieved by \ac{tof}-based transceivers.

Moreover, we investigate and propose a solution to the problem of ambiguous estimations affecting \ac{cots} transceivers, explained in Subsection~\ref{subsec:ambiguous-estimations}. We opt to focus our analysis on these transceivers for~\ref{enum:reasons-cots} main reasons:

\begin{enumerate}
	\item Signal-strength-based \ac{de} may lead to the development of simpler UWB transceivers featuring lower costs than those using \ac{tof}, which require high sampling rates~\cite{low-complexity-uwb}. These can be useful in applications affording accuracies up to a few decimeters. Throughout our experiments, we opted for sticking to the IEEE 802.15.4-2011 standard~\cite{standard-802154-2015} for compatibility purposes aiming to create alternatives which could serve as an extension to the existing standard rather than creating an incompatible approach, e.g., occupying the entire spectrum reserved for this technology.
	
	\item  We recently showed that the \textit{Distance Enlargement Fraud} - a particular attack on \ac{de} which cannot be overcome solely by using \ac{tof} measurements - can be detected or limited by using a novel framework relying on hybrid \ac{tof} and \ac{rss} distance estimations~\cite{botler-hybrid-tof-rss}. In this attack, a malicious entity $P$ tries to convince another entity performing \ac{de} to it that they are further away than they really are. While in a \ac{tof}-based system $P$ performs the attack by inserting a time delay in the response time, in a \ac{rss}-based system $P$ can amplify signals or communicate different power levels than those received, making this attack challenging to overcome. Our approach imposes bounds to those time delays and power gains by checking a set of geometrical constraints. As \ac{uwb} transceivers using \ac{tof} currently achieve a decimeter-level accuracy, the practicality of our approach is limited by the accuracy of \ac{rss} distance estimations obtained when using standard-compliant \ac{uwb} radios. 
	
	\label{enum:reasons-cots}
\end{enumerate} 


The contributions of this paper are:
\begin{itemize}
	\item We propose and evaluate a method to resolve ambiguities affecting \ac{dess} on \ac{cots} \ac{uwb} transceivers. The effectiveness of the proposed method is supported by extensive experiments using \ac{cots} devices, detailed in Section~\ref{sec:experiments}. 
	We have not found any publicly available dataset using the same set of parameters and features. In order to support future research in this topic, we make our datasets publicly available;
	\item We propose and analyze the suitability of several (54 in total) \ac{ml} regressors to improve the accuracy of \ac{uwb}-based \ac{dess}. The best model found achieved an accuracy as low as \SI{24}{\cm} in unknown environments, more than doubling the state-of-the-art accuracy~\cite{low-complexity-uwb}. 
\end{itemize}

%% file: sections/2-approach-and-methods.tex
\section{APPROACH AND METHODS}
\label{sec:approach-and-methods}

In this section, we explain the ambiguity issue affecting \ac{dess} on \ac{cots} \ac{uwb} transceivers, as well as our choices for features and how \ac{uwb} and \ac{ml} technologies are utilized.

\subsection{Background on \ac{uwb} Technology}
\label{subsec:background-uwb}

Due to the short pulse duration (\SI{\approx2}{ns}), \ac{uwb} technology enables the receiver to separate in time the signal received through the first path from the multipath reflected signals. A channel estimation, also known as \ac{cir}, is used by \ac{uwb} receivers to accurately determine the time point a transmitted pulse first reaches the receiver. To this end, a leading edge detection algorithm is typically applied on the absolute value of the \ac{cir}, whose samples are proportional to the power of the received signal, but, in \ac{cots} devices, are normalized, as will be discussed further in this paper. 

In our experiments, we use the DW1000~\cite{dw1000-datasheet} transceiver, which provides a \ac{cir} estimation by sampling the baseband received signal at a rate of \SI{\approx1}{GSPS}, and storing 1015 complex (1015 real + 1015 imaginary) \ac{cir} samples in memory. Those can be retrieved from the transceiver. 
Several examples of plots of absolute values of different \ac{cir}s can be seen in Figure~\ref{fig:cir-abs-plot}, where the X-axis' dimension is time, with a \SI{1}{\ns} interval between samples, and the Y-axis is proportional to the amplitude of the received signal.
We use only 32 out of the 1015 samples stored by the chip, as later samples were found to contain little power in the scenarios tested. Using this reduced amount of samples also reduces the complexity of our models. Note that the \ac{cir} itself contains no information about the time-of-arrival of the signal. The latter is stored in other registers, but are ignored in this paper.

\subsection{Resolving Ambiguous Estimations}
\label{subsec:ambiguous-estimations}
The DW1000 features an \ac{agc} stage in its frontend. \ac{agc}s are common in wireless receivers and enable longer communication ranges by amplifying the received signal. It aims to normalize the signal's amplitude by applying a low magnitude gain to high power signals and vice-versa. Thus, using the previously mentioned \ac{cir} samples without accounting for the gain applied by the \ac{agc} makes \ac{dess} difficult, if not meaningless. 

To assess the impact of the AGC on \ac{dess} accuracy, we perform one experiment, which is repeated twice in the same environment (a building hallway), initially with the \ac{agc} turned on, and then with it turned off. In both rounds, 2 \ac{cots} \ac{uwb} modules (TX and RX) were placed facing each other at different distances while TX transmits signals at different power levels to RX. 

\begin{figure}
    \includegraphics[width=\columnwidth]{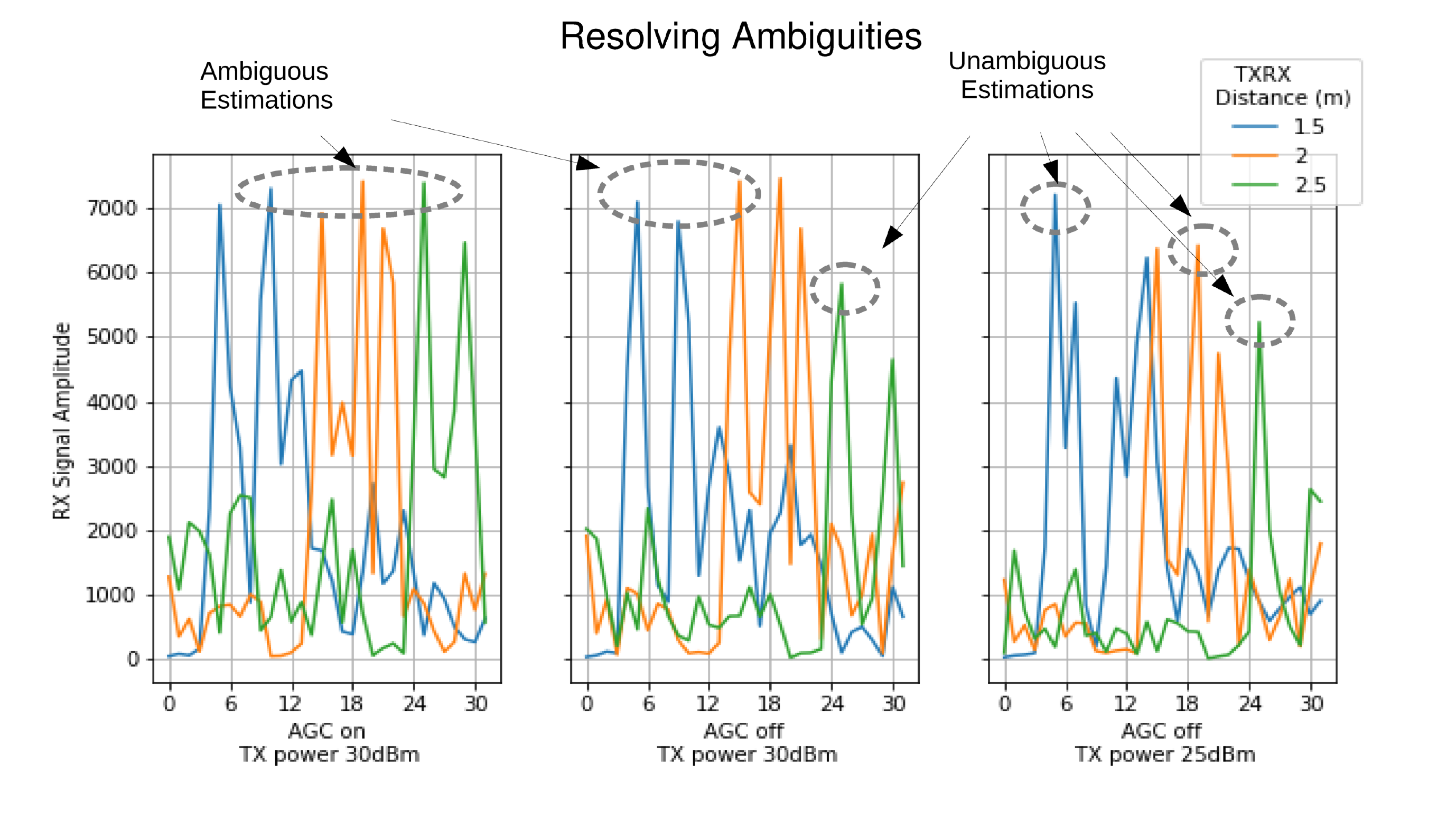}
    \caption{Plots of absolute values of selected CIRs samples at 3 different distances. The CIRs corresponding to TXRX distances of \SI{2}{\m} and \SI{2.5}{\m} are shifted in time (x-axis) to facilitate visualization. All received signals have the same amplitude in the left plot, which illustrates signals received using the \ac{agc} on at a high transmission power. In the middle plot, the furthest distance signal has a lower amplitude than the other, as a consequence of turning the \ac{agc} off. The rightmost plot, obtained with \ac{agc} off and a lower transmission power shows all the signals having different amplitudes.}
    \label{fig:cir-abs-plot}
    \vspace{-0.25cm}
\end{figure}

As shown in Figure~\ref{fig:cir-abs-plot} (middle), at shorter distances and high TX powers the RX saturates, leading to \textit{ambiguous} \ac{cir} estimations, i.e., a single CIR amplitude is associated to multiple TX-RX distances. This reduces the probability for the RX to correctly discriminate between distances. In this case, transmitting at lower power levels resolves the ambiguity issue as each CIR amplitude is associated to a unique distance, as shown on the right-most plot of the same figure. This experiment is analyzed in Subsection~\ref{subsec:agc-on-vs-agc-off}.  

\subsection{Features}
\label{subsec:features}
The set of features that we use is intentionally selected to reflect only the \ac{rss}. The features used are taken from~\cite{dw1000-user-manual} and listed in Table~\ref{tab:dw1000-acronyms}. They are all calculated by the transceiver and stored in registers, except for the absolute values of \ac{cir} samples whose complex values are stored.

\begin{table}[H]
    \begin{tabular}{|p{.2\columnwidth}|p{.6\columnwidth}|} \hline
        Abbreviation & Term\\ \hline
        FPPL         & First Path Power Level\\
        RSSI         & Received Signal Strength Indicator\\
        FP\_IDX      & First Path Index\\
        LDE\_PPAMPL  & Leading Edge Peak Path Amplitude\\
        LDE\_PPINDX  & Leading Edge Peak Path Index\\
        FP\_AMPLX    & First Path Amplitude Point X, X $\in \{1, 2, 3\}$ \\
        \hline
    \end{tabular}
    \caption{List of acronyms of features acquired.}
    \label{tab:dw1000-acronyms}
    \vspace{-0mm}
\end{table}

We use the FPPL metric to train and test our simplest models, discussed in Subsection~\ref{subsec:agc-on-vs-agc-off}. The FPPL is a scalar value calculated using 3 samples in the vicinity of the first peak path, and is, therefore, proportional to the power of this peak. Thus, the power from the reflected signals, which typically misleads power-based \ac{de}, has a less severe impact than when using other metrics, such as the RSSI. This result is presented in~\cite{low-complexity-uwb} and was confirmed by our experiments, but is not detailed in this paper.

In an attempt to improve the results obtained with the FPPL, we use the 32 absolute values of \ac{cir} samples as features, on top of which the other features can be estimated. The \ac{cir} contains information about the environment due to the reflected signals, which, in principle, do not interfere with the first path signal. Given that the data obtained in our experiments with \ac{agc} turned off leads to \ac{cir}s with amplitude values proportional to the received signal power - in such a way that the \ac{cir} contains information about the distance between the transceivers - we expect the models to learn how a \ac{cir} for a certain distance looks like. Whether or not these models generalize for different environments is still an open question, investigated in Subsections~\ref{subsec:using-cir-samples-as-features} to ~\ref{subsec:other-regressors}.

\subsection{\ac{ml}-Based Approach for \ac{de} and Regressors}
\label{subsec:ml-based-approach-and-regressors}

To train and test the \ac{ml} models we use data collected from the transceiver at known locations. The separation distance at each location serves as ground truth values. After the training phase, the \ac{ml} models are tested by estimating distances using the same features used for training, but different samples. 


As our target variable is continuous, we are interested in \textit{regression} models (opposed to classification models), which can provide as an output continuous \ac{de} values.
This enables estimating distances at a finer granularity than the \SI{0.5}{\m} distance step size used in our experiments - typically too coarse for \ac{uwb} \ac{de}.

Our analyses include several families of regressors classified according to~\cite{survey-regression-methods} as: linear and generalized linear models, LASSO and ridge regression, Bayesian models, Gaussian processes, nearest neighbors, regression trees and rules, random forests, bagging and boosting and support vector regression.




All \ac{ml} models are trained and tested in two different environments. While testing the models in the same environment where they have been trained provides an upper bound on the expected \ac{de} accuracy, testing them in a different environment enables us to check how well they generalize. Aiming to assess the information provided by the transmission power gain, we train k-Nearest Neighbors regressors in 2 different ways. First. we use as features the absolute values of the 32 CIR samples extended with the TX power gain, which is deterministic. Next, we remove the power gain from the feature set to compare with the results from the previous approach; the accuracy should vary according to the level of information provided by this feature.
These results should indicate whether to use the power gain as a feature, and are detailed in Subsections~\ref{subsec:using-cir-samples-as-features} and~\ref{subsec:discarding-power-gain}. This is essential in our next step - detailed in Subsection~\ref{subsec:other-regressors} - as we extend the analysis to other regressors, which may be more accurate and/or generalize better.
The performance of 54 out of the 55 regressors implemented in \textit{scikit-learn}~\cite{scikit-learn} version 1.0.2\footnote{Quantile Regressor was excluded from this analysis for requiring an extremely high training overhead.} is
 assessed using their default parameters.

%% file: sections/3-experiments.tex
\section{EXPERIMENTS}
\label{sec:experiments}

This section describes 3 experiments that we designed to assess the performance of UWB signal-strength-based \ac{de} with the goals of:
\begin{enumerate}[start=1,label={E.\Roman*)}]
    \item checking the influence of the \ac{agc} on signal-strength measurements obtained with the DW1000~\cite{dw1000-user-manual};
    \item testing the capability of estimating distances using only signal-strength-related features to train \ac{ml} models in a known environment. No \ac{tof}-related feature is utilized; 
    \label{enum:exp-hall}
    \item evaluating how the previously obtained models generalize, i.e., can perform \ac{de} in a different environment from the one it was trained in without any re-calibration. 
    \label{enum:exp-hallway}    
\end{enumerate}

\paragraph*{Hardware and Parameters}
For all our experiments we used as transmitter a DWM1003 module, which is an evaluation module from Decawave, embedding a single DW1000 chip~\cite{dw1000-datasheet}. As receiver, we used a DWM1002 module~\cite{igor-decawave}, which contains 2 DW1000 chips clocked by the same source, connected each to a dedicated antenna. The two antennas are separated by a distance of $\approx$\SI{2.05}{\cm}. Every packet transmitted was acquired by the two chips, and all packets correctly received by both chips were added to our dataset. For simplicity, we use data from a single receiver chip throughout our analysis. 

Standard channel 7 (center frequency=6489.6 MHz, bandwidth=1081.6 MHz) was used in all the experiments as well as clear \ac{los} between transceivers. All the 68 different programmable power gains $\in \{0~dB, 0.5~dB, ..., 33.5~dB\}$ available in the DW1000 were used, which we consider sufficient to provide a rich feature set. Other parameters were kept as default according to~\cite{uwb-core-mynewt}. 

\subsection{Experiment in a hallway with \ac{agc} turned on}
\label{subsec:exp-1}
This experiment was conducted in a $\approx$\SI{1.9}{\m} wide hallway. The modules were placed approximately \SI{1.5}{\m} above the floor and oriented along the length of the hallway with antennas facing each other. Their separation distance was varied from \SI{0.5}{\m} to \SI{6.5}{\m} in steps of \SI{0.5}{\m}, which was found to be the maximum communication distance - at maximum transmission power - when the receiver's \ac{agc} was turned off. Distances were measured with a measuring tape, so that errors in the range of centimeters are possible. For each distance, a minimum of 1088 ($=16*68$) packets were transmitted, in such a way that at least 16 packets were transmitted using each of the 68 power gains. The receiver's \ac{agc} was turned on. The features acquired are independent of ToF and include:
\begin{itemize}
    \item fppl, rssi, the fp\_idx, lde\_ppampl, lde\_ppindx, fp\_ampl1, fp\_ampl2 and fp\_ampl3. Please, refer to Table~\ref{tab:dw1000-acronyms} for a description of these acronyms and to \cite{dw1000-user-manual} for an explanation of their physical meaning;
    \label{item:features}
    \item 32 complex \ac{cir} samples, where the 5th sample corresponds to the first peak detected;
    \item the power gain value used by the transmitter. 
\end{itemize}

\subsection{Experiment in a hallway with \ac{agc} turned off}
\label{subsec:exp-2}
We repeat the experiments from Section~\ref{subsec:exp-1}, but with the receiver's AGC turned off. Please, notice that, in this experiment, many of the transmitted packets do not reach the receiver, depending on their power gain and on the communication distance.
Although it limits the communication range, eliminating the AGC enables us to simply establish an upper bound on the expected accuracy to be achieved using the proposed methods in case the gains provided by the \ac{agc} can be obtained or estimated.

\subsection{Experiment in a hall with \ac{agc} turned off}
\label{subsec:exp-3}
We repeat the experiments from Section~\ref{subsec:exp-2}, but in a wider ($\approx$\SI{9.3}{\m} x \SI{5}{\m}) building hall furnished only with working desks and chairs. The reason why the \ac{agc} was turned off in this experiment will be clarified in Subsection~\ref{subsec:agc-on-vs-agc-off}.

\subsection*{DATASETS}

Separate datasets were generated for each of the experiments~\ref{subsec:exp-1} to \ref{subsec:exp-3} and can currently be found on \cite{uwb-rss-dataset}, along with instructions on how to use them, as well as a description of the available features, which are not restricted to the ones used in our analysis.

%% file: sections/4-analysis-and-results.tex
\section{ANALYSIS AND RESULTS}
\label{sec:analysis-and-results} 

In this Section, we analyze the methods introduced in Section~\ref{sec:approach-and-methods} using the data previously obtained. The metric used to quantify accuracy is the \ac{mae} obtained per distance and then averaged over all distances, so that the final metric is not dominated by distances with greater sample sizes. MAE was preferred over \ac{rmse} as it equally weighs errors at different distances. Nonetheless, \ac{rmse} is occasionally used to enable direct comparisons with existing results using this metric. Additionally, our analyses include measures of memory and processing overhead for both training and testing the models.

\subsection{\ac{agc} On Vs \ac{agc} Off}
\label{subsec:agc-on-vs-agc-off}

\begin{figure}
    \includegraphics[width=\columnwidth]{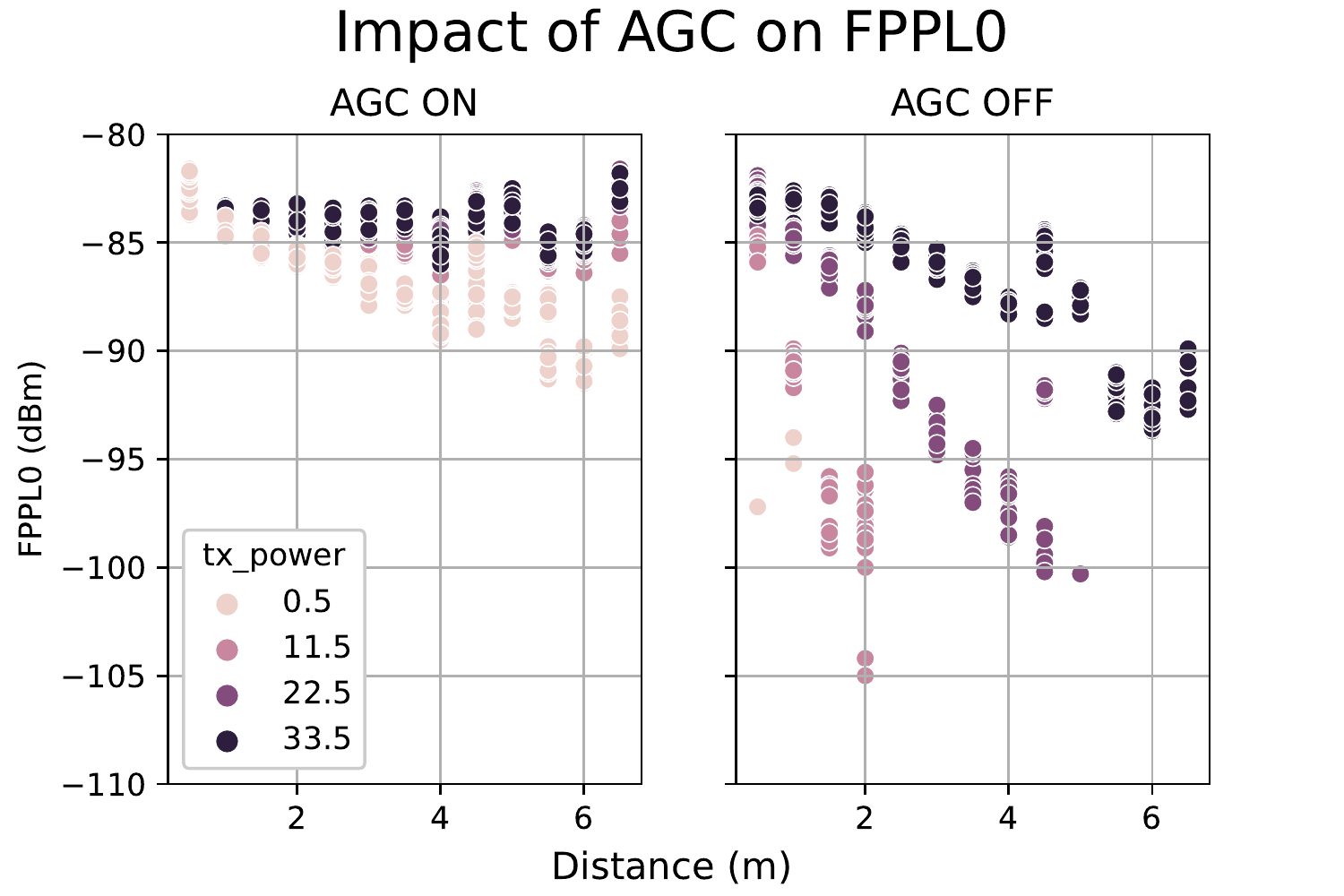}
    \caption{Scatter plot over distance of FPPL feature in the hallway with \ac{agc} turned on (left) and off (right) for 4 different transmission power gains.}
    \label{fig:scatter-agc-on-vs-agc-off}
\end{figure}

In order to achieve an optimum accuracy using only \ac{rss}-related features from the DW1000, we first evaluated how the \ac{agc} stage of the transceiver affects the accuracy of estimations.
Using data from experiments \ref{subsec:exp-1} and \ref{subsec:exp-2}, we show in Figure~\ref{fig:scatter-agc-on-vs-agc-off} a scatter plot of the FPPL feature over distance for different transmitted power levels. Please, recall that both experiments took place in the same environment, at the same fixed positions. 

From this figure, it is clear that, at the distance range observed, the impact of distance on FPPL is higher when the \ac{agc} is off. In other words, it is easier to distinguish among distances estimated using this metric due to a reduced overlap of samples obtained at different distances at a given power gain. In order to quantify this result, we evaluate the MAE of the K-Nearest Neighbors (KNN) regressor using the default parameters from \cite{scikit-learn} version 1.0.2. We used \SI{75}{\%} and \SI{25}{\%} of the data for training and testing, respectively. When using only the maximum transmission power, the averaged MAE decreased from \SI{1.421}{\m} when the \ac{agc} is turned on to \SI{0.413}{\m} when it is turned off, showing the benefit of turning the AGC off due to the effect illustrated in Figure~\ref{fig:cir-abs-plot} (middle). To further reduce ambiguities, we repeat the previous analysis over all transmission powers tested, resulting in a MAE reduction from \SI{1.282}{\m} when the \ac{agc} is turned on to \SI{0.190}{\m} when it is turned off, as illustrated in Figure~\ref{fig:cir-abs-plot} (right). The accuracy is improved by more than \SI{1}{\m}.

This result can be justified as 1) the role of the \ac{agc} is to mitigate the effect of power attenuation over distance, which counteracts the physical principle explored, and 2) in the vicinity of the receiver, the high gain provided by the \ac{agc} combined with high TX powers saturates the received signal, generating ambiguous measurements, as shown in Figure~\ref{fig:cir-abs-plot} (left). This effect is attenuated when the \ac{agc} is turned off, as shown in Figure~\ref{fig:cir-abs-plot} (middle). Finally, the remaining ambiguities - due to saturation of the receiver even when the AGC is off - can be mitigated by using lower power gains (right).

Having demonstrated the advantages of having the \ac{agc} turned off, we utilize only datasets obtained with \ac{agc} off in the remaining analyses. Those datasets stem from experiments~\ref{subsec:exp-2} and~\ref{subsec:exp-3}. 

It is important to highlight that the mean standard deviation of the FPPL feature over all samples obtained in the hall grouped by distance and transmit power was $0.636 dB$, which is very close from the value (of $0.64 dB$) reported in~\cite{low-complexity-uwb}. Therefore, the current approach alone does not improve \ac{de} accuracy comparing with the state-of-the-art approach. In fact, it is intended to enable \ac{dess} on \ac{cots}.

\subsection{Using CIR Samples as Features}
\label{subsec:using-cir-samples-as-features}

Further CIR peaks tend to be more attenuated than the first one (proportional to the FPPL), which can provide additional information to the models. Therefore, we evaluate the performance of using as feature set the absolute values of the 32 complex CIR samples extended by the power gain. We proceed as in Subsection~\ref{subsec:agc-on-vs-agc-off}\footnote{From now on, our models use only 2 neighbors and weight points based on the inverse of the distance. All features besides the power gain are standardized by removing the mean and scaling to unity variance.}, observing that we also vary the training and testing environment to assess the robustness of the models against multipath interference, i.e., how each model generalizes. The results are summarized in Table~\ref{tab:maes-cir32}. 

\begin{table}[]
    \begin{tabular}{|p{.4\columnwidth}|p{.275\columnwidth}|p{.175\columnwidth}|} \hline
        \diagbox{Training Set}{Test Set}   & {\centering Hall}           & {\centering Hallway} \\ \hline \hline
        \multirow{1}{*}{Hall}                   & 0.032                       & 0.512               \\ \hline
        \multirow{1}{*}{Hallway}                & 0.449                       & 0.045                \\ \hline
    \end{tabular}
    \caption{Average MAE values in meters when training a model and testing in 2 different environments, including the power gain in the feature set.}
    \label{tab:maes-cir32}
    \vspace{-10mm}
\end{table}

We see that when the training and testing environments are the same, the average MAE is limited to a few centimeters, while when testing the model in a different environment than the one it was trained in, the MAE is in the order of half a meter. Therefore, it is possible to recommend such an approach for fingerprinting-based applications requiring high accuracy. This first result already improves the one from Subsection~\ref{subsec:agc-on-vs-agc-off} with \ac{agc} off by \SI{\approx16}{\cm}. Furthermore, we can conclude that environments with different multipath characteristics require a training phase, i.e., the proposed model does not generalize well. For future reference, we refer to the current approach as the \textit{standard approach}. The sub-decimeter accuracy obtained is a consequence of overfitting, as the \ac{cir}s are stable and sensitive to the environment. 

Testing in the hall the FPPL models from Subsection~\ref{subsec:agc-on-vs-agc-off} - which were trained in the hallway - we get an average \ac{mae} of \SI{\approx0.28}{\m}, which is even better than the results obtained in this section. Furthermore, as the FPPL may saturate at lower distances and/or higher power gains, we should expect lower power gains to reduce ambiguity and further improve accuracy. In fact, when using only the lowest power gain in communication range with each distance in the training set, we improve the average \ac{mae} to \SI{\approx0.24}{\m}. This result more than doubles the accuracy achieved by the state-of-the-art approach (of \SI{0.52}{\m}). To find the minimum power gain, we consider transmitting a sounding packet at a high power gain, enabling a rough \ac{de}, followed by a second transmission at the coarsely estimated minimum power gain. 

It is important to mention that the accurate results from Table~\ref{tab:maes-cir32} were obtained when testing samples at distances already \textit{seen} by the model. If we remove all the samples at a given distance $d_i$ from the training phase, train the model with all the distanced $d_j \neq d_i$, and test samples only at this particular \textit{unseen} distance $d_i$, repeating the procedure for every distance in our experiment, we obtain an average MAE of \SI{0.642}{\m} in the hall and \SI{0.730}{\m} in the hallway, meaning that the accuracy depends on the positions seen by the model and drastically deteriorates when the receiver is located at a position different than those at which the model has been trained. This procedure is known as \enquote{k-Fold cross-validation with non-overlapping groups}.

The time spent to train this model using a total of 11318 samples was measured to be \SI{2.392}{\ms} while the time spent to predict the entire test set containing 3773 samples was \SI{2.400}{\s} resulting in an average prediction time per sample of \SI{\approx636}{\micro s}. Those measurements were executed using libraries already mentioned on an Intel® Core™ i7-9850H CPU @ 2.60GHz running Microsoft Windows Version 10.0.19042 Build 19042.
The models obtained in the hall and in the hallway occupy \SI{3007}{\kilo bytes} and \SI{1815}{\kilo bytes} in memory, respectively.


\subsection{Discarding Power Gain}
\label{subsec:discarding-power-gain}
In the previous subsection, we included the power gain used by the transmitter as a feature to train and test the obtained models without showing if there is a benefit of doing it. In this subsection, we remove it from the feature set. As a matter of fact, not transmitting this information reduces the complexity of the \ac{de} protocol. Table~\ref{tab:maes-cir32-no-power-gain} summarizes the results.

\begin{table}[h]
    \begin{tabular}{|p{.4\columnwidth}|p{.275\columnwidth}|p{.175\columnwidth}|} \hline
        \diagbox{Training Set}{Test Set}   & {\centering Hall}           & {\centering Hallway} \\ \hline \hline
        \multirow{1}{*}{Hall}                   & 0.051                       & 1.058                \\ \hline
        \multirow{1}{*}{Hallway}                & 0.741                       & 0.057                \\ \hline
    \end{tabular}
    \caption{Average MAE values in meters when ignoring the knowledge of the transmission power gain.}
    \label{tab:maes-cir32-no-power-gain}
    \vspace{-5mm}
\end{table}

As expected, all MAE's increased, specially when testing the model in a different environment than the one it was trained in. This result indicates that the knowledge of the transmit power is beneficial (if not critical) for approaches relying on different transmit powers. To our knowledge, this is the first time that such an approach is proposed for UWB. However, this feature is less critical when training and testing the models in the same environment, as the errors are still within decimeter range.

\subsection{Other Regressors}
\label{subsec:other-regressors}
Finally, the \textit{standard approach} was tested using other regressors. 
We show in Table~\ref{tab:best-regressors} the best MAEs achieved and the respective regressor for each combination of training and testing set derived from the two different environments. 
Values may differ from those of Table~\ref{tab:maes-cir32}, as we present here the overall MAE instead of the average MAE, for simplicity.

\begin{table}[]
    \begin{tabular}{|p{.4\columnwidth}|p{.275\columnwidth}|p{.175\columnwidth}|} \hline
        \diagbox{Training Set}{Test Set}   & {\centering Hall}           & {\centering Hallway} \\ \hline \hline
        \multirow{3}{*}{Hall}                   & KNeighbors                  & MLP                  \\
                                                & Regressor                   & Regressor            \\ 
                                                & 0.041                       & 0.804                \\\hline
        \multirow{3}{*}{Hallway}                & Orthogonal                  & KNeighbors           \\ 
                                                & MatchingPursuitCV           & Regressor            \\
                                                & 0.787                       & 0.021                \\\hline
    \end{tabular}
    \caption{Lowest MAE's and respective regressors for different training and testing environments.}
    \label{tab:best-regressors}
    \vspace{-10mm}
\end{table}

KNeighbors shows to be the best regressor when training and testing the model in the same environment. Otherwise, it can be outperformed even without tuning of parameters.

%% file: sections/5-related-work.tex

\section{RELATED WORK}
\label{sec:related-work}


Two modulation schemes are defined for UWB: pulse-based modulation, which targets low data rate applications, and Orthogonal Frequency Division Modulation (OFDM), which targets high data rate applications.
In \cite{rssi-uwb-ofdm} and~\cite{rssi-uwb-hrp}, the authors evaluated the accuracy achieved with RSSI for OFDM UWB (high data rate). This is not the most obvious approach, as multipath fading should be mostly mitigated when using short pulses in time, which is achieved with the pulse-based modulation. The authors characterized channel 15 based on RSSI measurements over a distance of only \SI{2}{\m} and reported mean \textit{positioning} errors in the range of \SI{10}{\cm} to \SI{30}{\cm}. Similarly, we include analyses using the first path power level (FPPL) instead of the RSSI, and we use \ac{cots} transceivers supporting pulse-based modulation. We present expected errors for \textit{\ac{de}} rather than for positioning. The latter depends on the number of anchors (devices at known positions) used, as well as on the used positioning algorithms.

In \cite{low-complexity-uwb}, the authors made use of UWB signal statistics for estimating distances. The accuracy of the approach is determined by the standard deviation of an - empirically obtained - Gaussian noise term in the log normal path-loss equation. The best accuracy is achieved under \ac{los} and equals \SI{0.52}{\m}, which, as shown in Section~\ref{sec:analysis-and-results}, could be improved down to \SI{0.021}{\m} when using our approach in a known environment, and down to \SI{0.24}{\m} even in unknown environments. Furthermore, the authors used in their experiments a Gaussian pulse generator with a \SI{20}{\ps} duration, which is much shorter than the $\approx$\SI{2}{\ns} standard compliant pulses, used in our experiments. Similarly, in \cite{uwb-channel-measurements} and \cite{gigl}, the authors showed path loss curves from experiments using the same setup as above. The former focused on modeling the UWB channel while the latter evaluated \textit{positioning} errors. 

In \cite{uwb-model-vna}, the \ac{uwb} channel is characterized using a vector network analyzer (VNA) occupying the whole \SI{7.5}{\GHz} UWB FCC spectrum. Likewise, a VNA was used in \cite{empirical-path-loss-2-pages} for performing UWB measurements in 4 different environments, with bandwidths of \SI{500}{\MHz} and \SI{3}{\GHz}. Both obtained parameters for a log normal path-loss model, but did not provide an approach to improve \ac{de} accuracy.



\ac{ml} has been recently proposed to improve distance and position estimations. In \cite{ml-approaches-for-network-loc-jku}, it was used to improve ToF distance estimations down to \SI{1.6}{\cm} \ac{rmse} in LOS conditions, using a set of features from 3 packets composing a dual-sided two-way ranging (DS-TWR) exchange. Our method requires a single packet instead and no \ac{tof}-related information. 

Many other approaches have been proposed aiming to correct ToF-based ranging estimations, such as~\cite{distance-measurements-feedforward-nn}. To our knowledge, the use of \ac{ml} methods has not been used to enable \ac{rss}-based \ac{uwb} \ac{de}, which is the gap covered in this paper.


%% file: sections/6-conclusion.tex
\section{CONCLUSIONS AND FUTURE WORK}
\label{conclusions}
In this paper we investigated the use of \ac{ml} regressors for \ac{uwb} \ac{de} using only signal-strength-related features and issues associated to its implementation on \ac{cots} transceivers. Successive analyses enabled finding more suitable parameters, features and regressors to this end. 

By using multiple transmit powers, we managed to achieve an accuracy of \textbf{24 cm} in an \textit{unknown} environment. This result \textbf{more than doubles} the accuracy achieved by the state-of-the-art approach. The best accuracy achieved with the methods proposed in a \textit{known} environment is as high as \textbf{\SI{2.1}{\cm}}.

Future work includes preprocessing the received signal, using different \ac{ml} models and tuning the models' parameters. Increasing the communication range of the proposed approaches by using the \ac{agc} gain as a parameter instead of turning it off may also be an interesting research direction. We also believe that conducting more experiments, for instance, subject to obstructed \ac{los} between transceivers, would encourage and enable the search towards better \ac{de} methods.

%% file: sections/8-acknowledgments.tex
\section*{ACKNOWLEDGEMENTS}
\label{acknowledgments}

This work has been supported by the FFG, Contract No. 881844: "Pro²Future" and the TU Graz LEAD project Dependable Internet of Things.

%% file: main.bbl
\begin{thebibliography}{10}
\providecommand{\url}[1]{#1}
\csname url@rmstyle\endcsname
\providecommand{\newblock}{\relax}
\providecommand{\bibinfo}[2]{#2}
\providecommand\BIBentrySTDinterwordspacing{\spaceskip=0pt\relax}
\providecommand\BIBentryALTinterwordstretchfactor{4}
\providecommand\BIBentryALTinterwordspacing{\spaceskip=\fontdimen2\font plus
\BIBentryALTinterwordstretchfactor\fontdimen3\font minus
  \fontdimen4\font\relax}
\providecommand\BIBforeignlanguage[2]{{%
\expandafter\ifx\csname l@#1\endcsname\relax
\typeout{** WARNING: IEEEtran.bst: No hyphenation pattern has been}%
\typeout{** loaded for the language `#1'. Using the pattern for}%
\typeout{** the default language instead.}%
\else
\language=\csname l@#1\endcsname
\fi
#2}}

\bibitem{dw1000-datasheet}
Decawave, \emph{\BIBforeignlanguage{English}{{DW1000} {IEEE802.15.4-2011} {UWB}
  Transceiver datasheet}}, Decawave Ltd, 2015.

\bibitem{low-complexity-uwb}
G.~Bellusci, G.~J.~M. Janssen, J.~Yan, and C.~C. J.~M. Tiberius, ``Low
  complexity ultra-wideband ranging in indoor multipath environments,'' in
  \emph{2008 IEEE/ION Position, Location and Navigation Symposium}, 2008, pp.
  394--401.

\bibitem{standard-802154-2015}
``{IEEE} standard for low-rate wireless networks,'' The Institute of Electrical
  and Electronics Engineers, Inc., New York, USA, Standard, dec 2015.

\bibitem{botler-hybrid-tof-rss}
L.~Botler, K.~Diwold, and K.~R{\"o}mer, ``\BIBforeignlanguage{English}{A
  {UWB}-based solution to the distance enlargement fraud using hybrid {ToF} and
  {RSS} measurements},'' in \emph{\BIBforeignlanguage{English}{Proceedings -
  2021 IEEE 18th International Conference on Mobile Ad Hoc and Smart Systems,
  MASS 2021}}.\hskip 1em plus 0.5em minus 0.4em\relax IEEE Publications, Oct.
  2021, pp. 324--334.

\bibitem{dw1000-user-manual}
Decawave, \emph{\BIBforeignlanguage{English}{DW1000 User Manual}}, Decawave
  Ltd, 2017.

\bibitem{survey-regression-methods}
M.~Fern{\'a}ndez-Delgado, M.~S. Sirsat, E.~Cernadas, S.~Alawadi, S.~Barro, and
  M.~Febrero-Bande, ``An extensive experimental survey of regression methods,''
  \emph{Neural Networks}, vol. 111, pp. 11--34, 2019.

\bibitem{scikit-learn}
F.~Pedregosa, G.~Varoquaux, A.~Gramfort, V.~Michel, B.~Thirion, O.~Grisel,
  M.~Blondel, P.~Prettenhofer, R.~Weiss, V.~Dubourg, J.~Vanderplas, A.~Passos,
  D.~Cournapeau, M.~Brucher, M.~Perrot, and E.~Duchesnay, ``Scikit-learn:
  Machine learning in {P}ython,'' \emph{Journal of Machine Learning Research},
  vol.~12, pp. 2825--2830, 2011.

\bibitem{igor-decawave}
I.~{Dotlic}, A.~{Connell}, H.~{Ma}, J.~{Clancy}, and M.~{McLaughlin}, ``Angle
  of arrival estimation using decawave {DW1000} integrated circuits,'' in
  \emph{2017 14th Workshop on Positioning, Navigation and Communications
  (WPNC)}, 2017, pp. 1--6.

\bibitem{uwb-core-mynewt}
``{UWB}-core,'' \url{https://github.com/decawave/uwb-core}, 2022, accessed:
  2022-03-24.

\bibitem{uwb-rss-dataset}
L.~Botler, ``{UWB} {RSS} {Dataset},''
  \url{https://bitbucket.org/leobotler/uwb-rss-dataset}, accessed: 2022-04-19.

\bibitem{rssi-uwb-ofdm}
A.~{Waadt}, A.~{Burnic}, D.~{Xu}, C.~{Kocks}, S.~{Wang}, and P.~{Jung},
  ``Analysis of {RSSI} based positioning with multiband {OFDM} {UWB},'' in
  \emph{2010 Future Network Mobile Summit}, 2010, pp. 1--8.

\bibitem{rssi-uwb-hrp}
S.~{Wang}, A.~{Waadt}, A.~{Burnic}, D.~{Xu}, C.~{Kocks}, G.~H. {Bruck}, and
  P.~{Jung}, ``System implementation study on rssi based positioning in {UWB}
  networks,'' in \emph{2010 7th International Symposium on Wireless
  Communication Systems}, 2010, pp. 36--40.

\bibitem{uwb-channel-measurements}
Z.~Irahhauten, G.~J. Janssen, H.~Nikookar, A.~Yarovoy, and L.~P. Ligthart,
  ``{UWB} channel measurements and results for office and industrial
  environments,'' in \emph{2006 IEEE International Conference on
  Ultra-Wideband}, 2006, pp. 225--230.

\bibitem{gigl}
T.~Gigl, G.~J. Janssen, V.~Dizdarevic, K.~Witrisal, and Z.~Irahhauten,
  ``Analysis of a {UWB} indoor positioning system based on received signal
  strength,'' in \emph{2007 4th Workshop on Positioning, Navigation and
  Communication}, 2007, pp. 97--101.

\bibitem{uwb-model-vna}
\BIBentryALTinterwordspacing
L.~Rubio, J.~Reig, H.~Fern{\'a}ndez, and V.~M. Rodrigo-Pe{\~{n}}arrocha,
  ``Experimental {UWB} propagation channel path loss and time-dispersion
  characterization in a laboratory environment,'' \emph{International Journal
  of Antennas and Propagation}, vol. 2013, p. 350167, Mar 2013. [Online].
  Available: \url{https://doi.org/10.1155/2013/350167}
\BIBentrySTDinterwordspacing

\bibitem{empirical-path-loss-2-pages}
N.~Alsindi, B.~Alavi, and K.~Pahlavan, ``Empirical pathloss model for indoor
  geolocation using {UWB} measurements,'' \emph{Electronics Letters}, vol.~43,
  no.~7, pp. 370--372, 2007.

\bibitem{ml-approaches-for-network-loc-jku}
J.~W. Karoliny, ``Machine learning approaches for high accuracy network
  localization with {UWB}/submitted by julian karoliny,'' Ph.D. dissertation,
  Universit{\"a}t Linz, 2020.

\bibitem{distance-measurements-feedforward-nn}
\BIBentryALTinterwordspacing
P.~Krapež, M.~Vidmar, and M.~Munih, ``Distance measurements in {UWB}-radio
  localization systems corrected with a feedforward neural network model,''
  \emph{Sensors}, vol.~21, no.~7, 2021. [Online]. Available:
  \url{https://www.mdpi.com/1424-8220/21/7/2294}
\BIBentrySTDinterwordspacing

\end{thebibliography}
